\documentclass[prl,aps,twocolumn,superscriptaddress,showpacs,nofootinbib]{revtex4-2}
\usepackage{graphicx}
\usepackage{dcolumn}
\usepackage{bm}
\usepackage{gensymb}
\usepackage{multirow}
\usepackage{amsmath,amsthm,array}
\usepackage{amsfonts}
\usepackage{booktabs}
\usepackage{siunitx}
\usepackage{soul}
\usepackage{hyperref} 

\newcommand{\CVSS}{CsV$_{3}$Sb$_{5-x}$Sn$_{x}$}
\newcommand{\AVS}{\textit{A}V$_{3}$Sb$_{5}$}
\newcommand{\CVS}{CsV$_{3}$Sb$_{5}$}

\begin{document}

\preprint{APS/123-QED}

\title{\textbf{Interplay between charge correlations and superconductivity across the superconducting domes of CsV$_{3}$Sb$_{5-x}$Sn$_x$ 
}}%

\author{Andrea N. Capa Salinas}
\affiliation{Materials Department, University of California, Santa Barbara, CA 93106, USA}

\author{Brenden R. Ortiz}
\affiliation{Materials Science and Technology Division, Oak Ridge National Laboratory, Oak Ridge, TN 37831, USA}

\author{Steven J. Gomez Alvarado}
\affiliation{Materials Department, University of California, Santa Barbara, CA 93106, USA}

\author{Sarah Schwarz}
\affiliation{Physics Department, University of California, Santa Barbara, CA 93106, USA}

\author{Ganesh~Pokharel}
\affiliation{Materials Department, University of California, Santa Barbara, CA 93106, USA}
\affiliation{Perry College of Sciences, University of West Georgia, Carrollton, GA 30118, USA}

\author{Luca Buiarelli}
\affiliation{Department of Chemical Engineering and Materials Science, University of Minnesota, Twin Cities, Minneapolis, MN 55455, USA}

 \author{Hyeonseo Harry Park}
\affiliation{Department of Chemical Engineering and Materials Science, University of Minnesota, Twin Cities, Minneapolis, MN 55455, USA}

\author{Shiyu Yuan}
\affiliation{Department of Creative Studies, University of California, Santa Barbara, CA 93106, USA}

\author{Roland Yin}
\affiliation{Department of Electrical \& Computer Engineering, University of California, Santa Barbara, CA 93106, USA}
\affiliation{Materials Department, University of California, Santa Barbara, CA 93106, USA}

\author{Suchismita Sarker}
\affiliation{Cornell High Energy Synchrotron Source, Cornell University, Ithaca, NY 14853, USA}

\author{Turan Birol}
\affiliation{Department of Chemical Engineering and Materials Science, University of Minnesota, Twin Cities, Minneapolis, MN 55455, USA}

\author{Stephen D. Wilson}
 \email[email: ]{stephendwilson@ucsb.edu}
\affiliation{Materials Department, University of California, Santa Barbara, CA 93106, USA}
 

\date{\today}

\begin{abstract}
 The kagome metal CsV$_3$Sb$_5$ shows an unconventional interplay between charge density wave (CDW) order and superconductivity.  Tuning the band filling is known to rapidly suppress long-range CDW order and drive the formation of two superconducting ``domes" upon increasing hole concentration. Here we determine the detailed evolution of charge correlations across this phase diagram and resolve their interplay with the superconducting state. Upon light hole-doping, the suppression of a metastable $2\times 2\times 4$ CDW state coincides with the suppression of superconducting fluctuations present in the parent CsV$_3$Sb$_5$ compound.  Continued doping suppresses long-range $2\times 2\times 2$ CDW order, leaving remnant short-range, quasi-1D correlations that persist across the second superconducting dome.  These higher temperature charge correlations are seemingly essential to the lower temperature superconducting state, as charge correlations vanish coincident with superconductivity as a function of hole-doping. A multidomain model of short-range V-V dimer formation within the kagome plane is proposed in the second superconducting dome, where rotational and translational symmetry remain locally broken even in the absence of long-range CDW order.  
\end{abstract}

\maketitle


The lattice geometry of kagome metals gives rise to the formation of compact localized states and flat bands via kinetic frustration \cite{PhysRevB.78.125104, PhysRevLett.106.236802, PhysRevLett.106.236804} as well as Van Hove singularities (VHS) that harbor sublattice interference effects \cite{PhysRevB.86.121105}.  Electronic interactions, when added to either flat band or saddle-point fillings in the kagome lattice, are non-perturbative and are predicted to give rise to a diverse range of emergent phenomena \cite{disante2025kagomemetals}. These include unconventional superconductivity \cite{PhysRevLett.110.126405, PhysRevB.87.115135}, electronic nematicity \cite{PhysRevB.107.155131}, orbital magnetism \cite{PhysRevLett.132.146501, PhysRevB.104.035142, zhan2025loopcurrentorderkagome}, and intertwined orders such as pair density wave instabilities \cite{PhysRevB.108.L081117}. 

The \AVS~($A$ = K, Rb, Cs) family of kagome metals possess a band filling close to pure-type (\textit{p}-type) Van Hove singularities associated with their vanadium-based kagome networks \cite{Wilson2024}.  The amplified relevance of nearest-neighbor Coulomb interactions combined with multiorbital effects are proposed to drive the formation of unconventional charge density wave (CDW) order that coexists with an anomalous superconducting state \cite{ortiz2021superconductivity, PhysRevLett.125.247002, yin2021superconductivity}. In particular, the real component of the CDW state seemingly forms alongside an ``imaginary" CDW or orbital antiferromagnetic state that breaks time-reversal symmetry \cite{xing2024optical, jiang2021unconventional, mielke2022time, guo2022switchable}. The lower temperature, singlet superconducting phase \cite{roppongi2023bulk,mu2021s,PhysRevB.110.144516,PhysRevB.108.144508} is then a candidate for a chiral $d + id$ pairing state, and the interplay between the higher temperature CDW order and the lower temperature superconducting (SC) state remains an area of active investigation.



In \CVS, the CDW state exists on the verge of metastability. A complex three-dimensional CDW structure is observed, composed of phase-separated $2 \times 2 \times 2$ and $2 \times 2 \times 4$ reconstructed cells whose relative volume fractions and stabilities can be biased via external perturbations such as thermal cycling, impurity content, and strain \cite{Plumb2024,PhysRevResearch.5.L012032}. Interestingly, this metastability is also accompanied by a unique electronic phase diagram tuned as a function of carrier filling and hydrostatic pressure \cite{PhysRevLett.126.247001, yu2021unusual, oey_fermi_2022}.   

Unlike RbV$_3$Sb$_5$ and KV$_3$Sb$_5$ \cite{PhysRevMaterials.6.074802}, when holes are doped into CsV$_3$Sb$_5$, two superconducting ``domes" form via a nonmonotonic evolution of the SC transition temperature ($T_c$) as long-range CDW order is suppressed. A rapid suppression of the $2 \times 2 \times 4$ CDW state is evident upon hole-doping and is accompanied by an increase in $T_c$ \cite{kautzsch2023incommensurate}.  Further doping suppresses $T_c$ along with long-range CDW order.  Remarkably, however, $T_c$ recovers beyond this phase boundary, and initial studies have shown that short-range charge correlations persist into the second ``dome".  How these charge correlations evolve upon hole-doping and their relation to SC remain open questions, and they are important for understanding the relevance of charge correlations to the underlying SC order parameter.

Here we determine the evolution of charge correlations and their interplay with SC in single crystals of \CVSS~across the hole-doping phase diagram.  Hole-doping drives a rapid suppression of the metastable CDW phase in \CVS, which is accompanied by a suppression of the strong SC fluctuations in the pristine compound.  Continued doping results in the formation of short-range charge correlations and the wave vector associated with these quasi-1D correlations approaches \textbf{q}=($\delta$, 0, $\frac{1}{2}$) with $\delta\approx\frac{3}{8}$ near optimal doping, consistent with \textit{ab initio} models of a relative enhancement of a lattice instability at the R-point. The diffuse charge correlations are modeled as local chains of V-V dimers that break rotational and translational symmetry and their disappearance with doping correlates with the suppression of the SC state. The implications of these findings relative to theoretical models of SC in CsV$_3$Sb$_5$ are discussed.

\begin{figure}[]
 \centering
 \includegraphics[width=1\columnwidth]{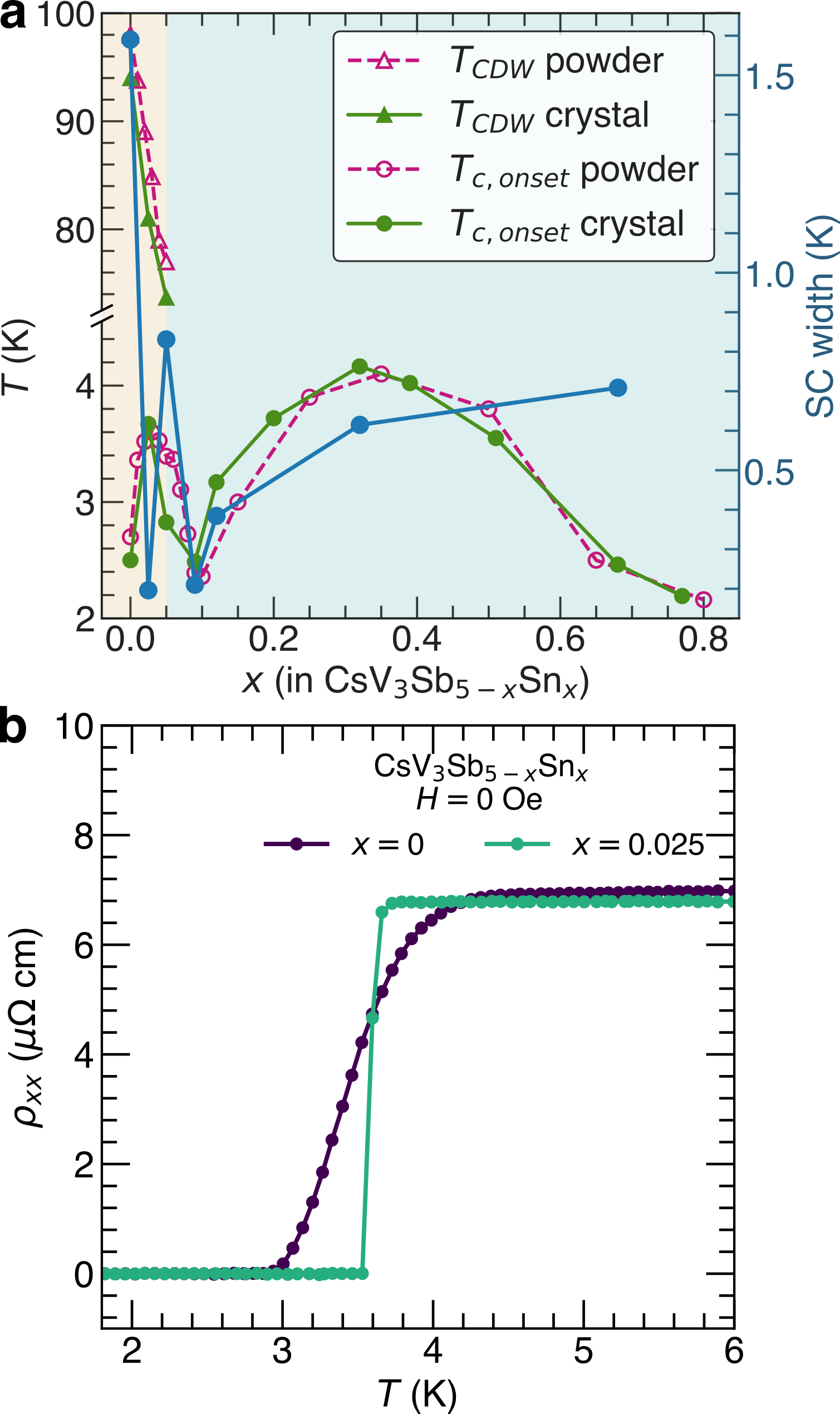}
 \caption{(a) Electronic phase diagram of CsV$_3$Sb$_{5-x}$Sn$_x$.  Closed symbols are from single crystals from this study and open symbols show polycrystalline data adapted from \cite{oey_fermi_2022}. The two phase diagrams match, illustrating the homogeneity of the crystals.  Right-hand axis shows the width of the SC transition as a function of hole doping $x$ (b) Two representative sets of electrical resistivity data for $x=0$ and $x=0.025$ crystals.}
 \label{fig:diagram}
\end{figure}

The electronic phase diagram showing the interplay between the long-range CDW order and SC is shown in Fig. 1 (a) for \CVSS~ crystals. Phase transitions were determined via magnetization measurements and were corroborated via resistivity data \cite{Supplemental}. The onset of CDW order is reported via the high-temperature peak in $ \partial   M/\partial   T$, and $T_c$ was determined via the onset of diamagnetism.  

The phase diagram shows a complicated interplay between the CDW and SC states. An initial suppression of the CDW state correlates with the increase in $T_{c}$, which reaches a local maximum at $x=0.025$. As doping increases further, a minima is observed at $x=0.09$, and, beyond this, $T_{c}$ recovers. $T_{c}$ reaches an absolute maximum near $x=0.32$, where $T_{c}=$~4.16~K, before then being suppressed with further doping.  The thermodynamic anomaly associated with the onset of CDW order is broadened and suppressed as a function of hole-doping and disappears near $x=0.05$; however this does not mean charge correlations vanish at this concentration.  Instead, as we will show later, it indicates a transition into a regime of short-range charge correlations.

One way of assessing the interplay between the SC transition and the evolution of charge correlations is to analyze the SC transition sampled via electrical transport.  The right-hand axis of Fig. 1 (a) shows the evolution of the width of the SC transition, defined as the thermal range between the onset of SC and the zero resistance state, as a function of hole-doping ($x$).  The SC transition of CsV$_3$Sb$_5$ has an anomalously broad transition, where $\rho (T)$ begins to decrease below $\approx 4.4$ K and a zero-resistance state is not reached until $T_c\approx 2.8$ K.  Given that the undoped system consistently has the broadest SC transition, this is unlikely to originate from trivial disorder and, instead, reflects strong SC fluctuations in the parent material.

Once a small amount of dopant is added, the zero-resistance $T_c$ increases, yet the SC transition sharpens dramatically.  This sharp change in the transition is illustrated in the $\rho(T)$ data shown in Figure 1 (b). Continued doping results in a broadening of the $T_c$ near $x=0.05$ near the phase boundary where long-range CDW vanishes.  This suggests a second phase boundary where fluctuation effects are enhanced at the border between SC domes.  Doping beyond this boundary results in $T_c$ sharpening before broadening again with increased doping/disorder.  These data establish two regimes of fluctuation effects in the SC order parameter, near $x=0$ and near $x=0.05$, that potentially connect to changes in the parent CDW state.  This is explored next via x-ray diffraction measurements.

\begin{figure}[]
 \centering
 \includegraphics[width=0.95\columnwidth]{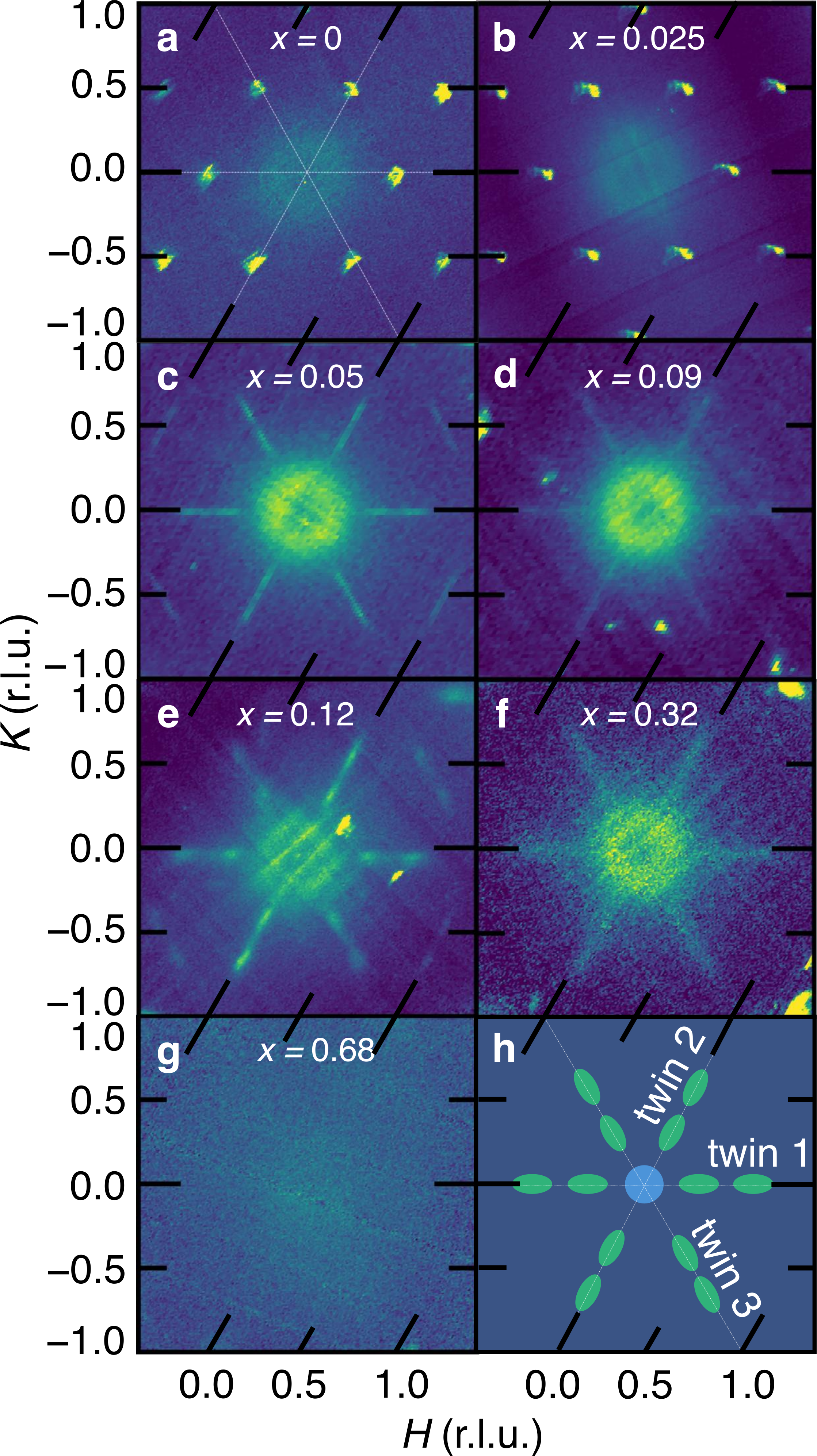}
 \caption{\label{fig:L-7p5}($H$, $K$)-maps of x-ray scattering data at a representative $L=7.5$ plane collected for a variety of hole concentrations $x$. Panels (a-f) show the evolution of in-plane charge correlations as hole-doping in increased from $x=0$ through $x=0.68$ where charge correlations vanish. Panel (h) shows a schematic of the quasi-1D charge correlations parametrized in a three-domain or three-twin model. }
\end{figure}

Synchrotron x-ray scattering measurements were performed across the phase diagram in Fig. 1 (a) with representative maps of scattering data shown in Fig. 2.  For $x=0$, Fig. 2 (a) shows sharp Bragg peaks at CDW superlattice positions corresponding to the $2\times 2$ enlargement of the $ab$-plane. Upon hole-doping, CDW order initially remains long-range in-plane; however the coexistence of $2\times 2\times 4$ and $2\times 2\times 2$ CDW states vanishes, leaving only the staggered TrH $2\times 2\times 2$ state \cite{Supplemental}.  This coincides with the rapid increase in the zero-resistance $T_c$ and the suppression of the extended SC fluctuations observed in the pristine system.

Upon doping to $x=0.05$, the charge correlations become short-range within the $ab$-plane.  This corresponds to the second position of a broadened $T_c$ in the phase diagram where the thermodynamic anomalies associated with long-range CDW order vanish. Diffuse correlations along $c$ remain peaked at $L=\frac{1}{2}$-type positions; however the in-plane correlations move away from the commensurate positions. The correlation volume of each peak also becomes anisotropic in-plane, with the scattering being more diffuse along the wave vector \textbf{q$_{\parallel}$} and sharper orthogonal to \textbf{q$_{\perp}$}.  Due to this anisotropy, a model of 120$^\circ$ rotated domains was adopted to parametrize the scattering (Fig. 2 (h)).  This represents a local breaking of the average 6-fold symmetry of the lattice.

Figs. 2 (c-g) show scattering maps charting the evolution of short-range CDW correlations from $x=0.05$ to $x=0.68$.  Short-range order persists throughout the second SC dome and vanishes below resolution at $x=0.68$.  This is the concentration where samples become partial volume fraction SCs \cite{oey_fermi_2022}. Note the central diffuse ``donut" shape at the zone center is an artifact from the diffuse tail of both neighboring primary Bragg peaks (out-of-plane).    

Figs. 3 (a-c) show the parametrization of charge correlations shown in Fig. 2. Each domain was characterized via cuts along \textbf{q$_{\parallel}$} and \textbf{q$_{\perp}$}.  At the $x=0.05$ doping level, charge correlations become quasi-1D, with a long ($\approx 160$ $\mathrm{\AA}$) correlation length along \textbf{q$_{\perp}$} and a short ($\approx 60$ $\mathrm{\AA}$), in-plane correlation length along \textbf{q$_{\parallel}$}. The wave vector \textbf{q}=($\delta$, 0, $\frac{1}{2}$) initially shifts away from the zone boundary ($\delta=\frac{1}{2}$), to $\delta=0.41$ (r.l.u.) at the boundary where long-range CDW order vanishes, and continued doping results in a shift toward $\delta=0.37$ (r.l.u.) near the peak of the second SC dome ($x=0.32$). Both \textbf{q$_{\perp}$} and \textbf{q$_{\parallel}$} correlation lengths shorten with increasing $x$, but the correlation volume remains qualitatively quasi-1D. Figs 3 (d-f) show representative directions, cuts, and fits to the data used to determine these parameters.



\begin{figure}[h]
 \centering
 \includegraphics[width=0.95\columnwidth]{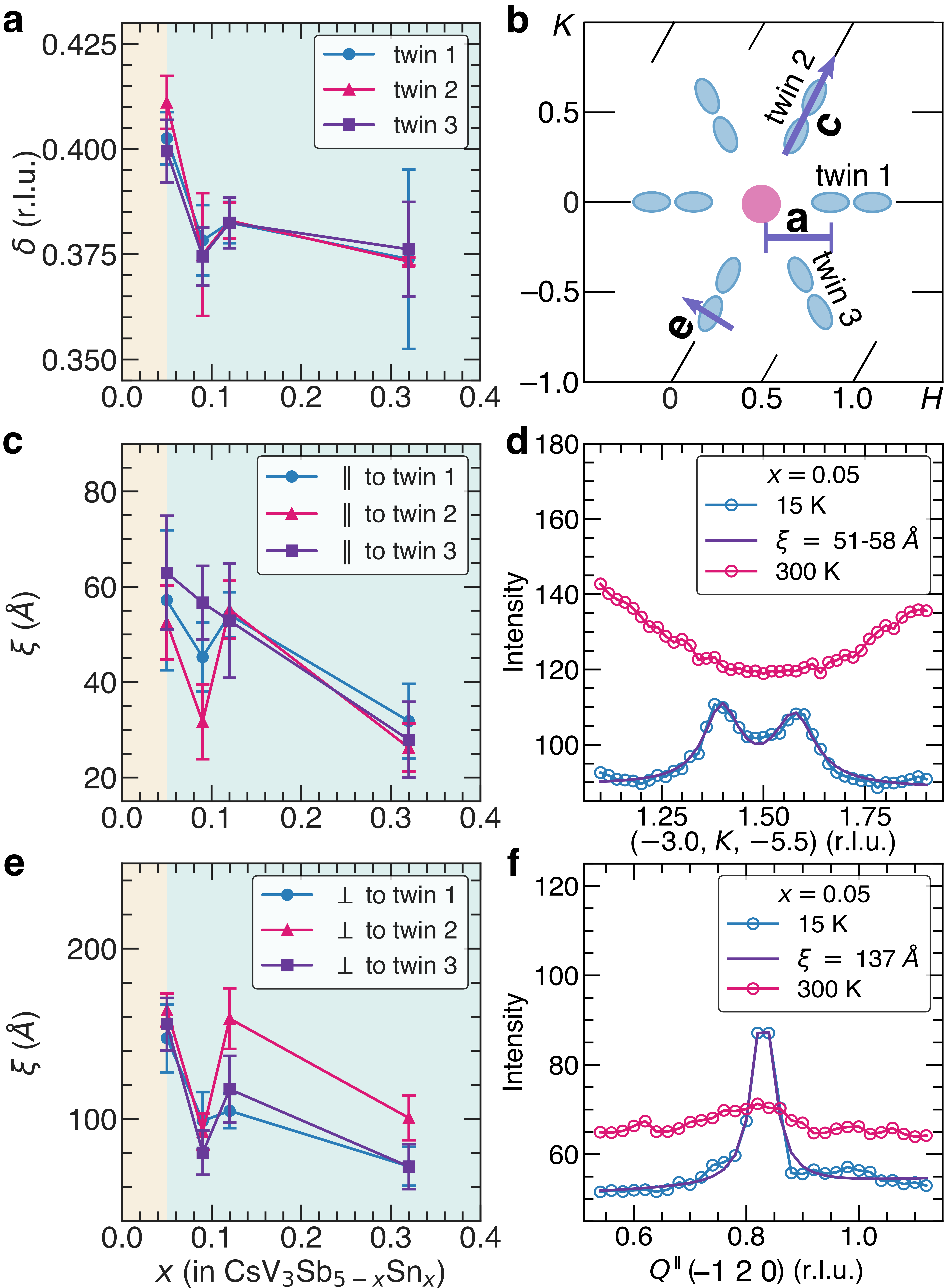}
 \caption{Parametrization of charge correlations shown in Fig. 2 (a) The evolution of the wave vector for the single-\textbf{q} state as it moves away from the L-point. (b) Tracking the correlation length parallel and perpendicular to a double domain show after $x=$0.09, the correlation length tends to decrease slightly, as shown in c,e. d and f portray how a cut is perform. Fitting is achieve with Lorentzian functions and a linear background.}
 \label{fig:Cuts}
\end{figure}


\begin{figure*}[]
 \centering
 \includegraphics[width=\textwidth]{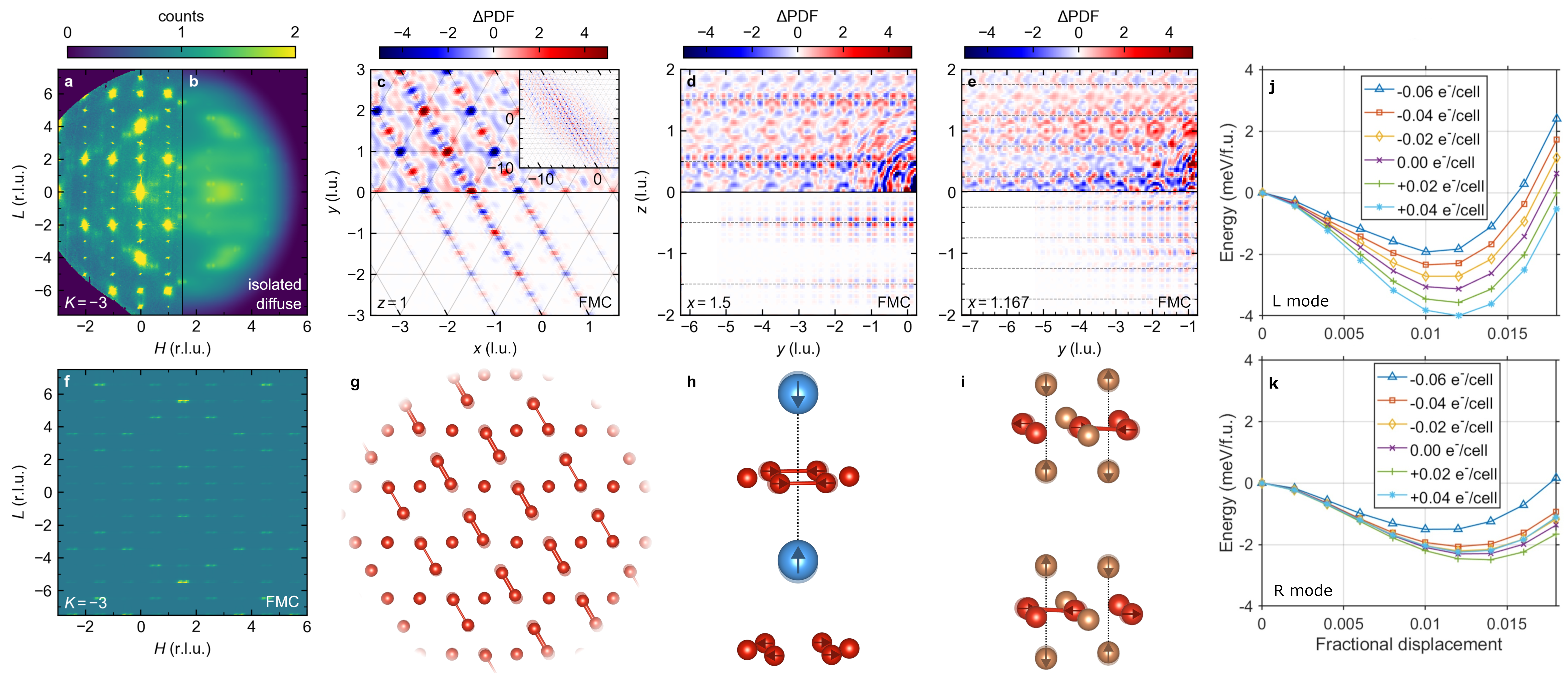}
 
\caption{(a) Total X-ray scattering and (b) isolated diffuse scattering in the $(H,-3,L)$ scattering plane. (c) Experimentally generated 3D-$\Delta$PDF and corresponding forward Monte Carlo results in the representative real space $(x,y,1)$ plane, (d) $(\frac{3}{2},y,z)$ plane, and (e) $(\frac{7}{6},y,z)$ plane. Inset of (c) presents longer range correlations from the experimentally generated 3D-$\Delta$PDF. (f) Calculated diffuse scattering in the $(H,-3,L)$ scattering plane obtained from forward Monte Carlo modeling. (g) Schematic of distortions employed in the forward Monte Carlo model including in-plane displacements of V, (h) out-of-plane displacements of Cs, and (i) out-of-plane displacements of Sb. Bonds represent V-V dimers, and their thickness is proportional to bond length.  (j) Energy from density functional theory as a function of L-mode amplitude, for different amounts of doping. For simplicity, the single-\textbf{q} displacement pattern is considered (k) Same as (j), but for the R-mode. }
 \label{fig:3DdPDF}
\end{figure*}

The short-range, quasi-1D correlations in the $L=\frac{1}{2}$-type scattering planes can be simulated using this three domain model where each domain includes correlations which are strongest along one of the equivalent $\langle 1,0,0 \rangle$, $\langle 0,1,0 \rangle$, $\langle 1,1,0 \rangle$ crystallographic directions. Three-dimensional difference pair distribution function (3D-$\Delta$PDF) transformation of the isolated diffuse scattering visualizes the variance of charge scattering in real space, which can then be modeled using Monte Carlo methods. 

Total scattering and isolated diffuse scattering for a representative $(H,L)$ plane are shown in Figures 4(a) and 4(b), respectively, where only a single domain of diffuse scattering is isolated from the data. A Fourier transform of the isolated diffuse scattering reveals strong quasi-1D displacive correlations along the $[0,1,0]$ direction which alternate in sign. This likely originates from a unidirectional, in-plane dimer instability within the V kagome network \cite{tazai2022mechanism}.  Along the perpendicular $[2,1,0]$ direction, the correlation length is significantly shorter, in agreement with the broader FWHM of the diffuse scattering along the \textbf{q$_{\parallel}$} in reciprocal space.


In addition to the in-plane V displacements, other notable 3D-$\Delta$PDF signatures appear at $\langle 1/2,1/2,1/2 \rangle$- and $\langle 1/6,1/3,1/4 \rangle$-type positions, corresponding to the interatomic vectors connecting V-Cs and V-Sb (out-of-plane). These correlations adopt a quadrupole-like shape in the $yz$-planes, illustrating a correlation between the in-plane displacement of the V sites along $[0,1,0]$ and an out-of-plane displacement of the Cs and out-of-plane Sb sites. 


To help interpret 3D-$\Delta$PDF features, a forward Monte Carlo model was built to capture a dimerization instability within the V kagome layer and correlated displacements of the Cs and Sb sites above and below these layers. The in-plane displacive correlations were well-modeled by dimers of V which exhibited strong correlations along $[1,0,0]$ but a periodic, incommensurate modulation of these displacements along the perpendicular $[2,1,0]$ direction [Fig. 4(c,g)]. In order to capture the correct sign of the out-of-plane correlations observed in the 3D-$\Delta$PDF, the nearest Cs site needs to displace towards any V hexagon which experienced a contraction (two dimers) and away from those which experienced an expansion (zero dimers) (Fig. 4(d,h)). Similarly, it was determined that the nearest out-of-plane Sb sites tend to displace away from any V triangle which experiences a contraction (one dimer) and towards any V triangle which experiences an expansion (zero dimers) (Fig. 4(e,i)). 


The calculated scattering is shown in Figure 4(f) and shows good agreement with intensity modulations in the experimental data, albeit with an artificially smaller FWHM along $L$ (long correlation length along $c$), which was necessary to simplify the simulation.
%
The relative amplitude of each displacement mode was estimated via comparison of the calculated and experimental scattering (Fig. 4). Final displacement amplitudes were chosen to be $|dy_\mathrm{V}| = 0.13737~\mathrm{\AA}$, $|dz_\mathrm{Cs}| = 0.05495~\mathrm{\AA}$, and $|dz_\mathrm{Sb}| = 0.02747~\mathrm{\AA}$.


The introduction of holes into CsV$_3$Sb$_5$ has a strong impact on the parent CDW state, which, in turn, impacts the low-temperature SC state.  Hole-doping rapidly lifts CDW phase competition and by $2\%$ holes/f.u. globally stabilizes the staggered TrH state, corresponding to a sharpening of the SC transition and demonstrating a link between CDW phase competition and the SC state. While the $T_c$ defined via the onset of zero resistivity shifts higher with doping, the initial onset of the downturn in resistivity actually shifts downward.  This suggests that absent fluctuations between CDW states, CsV$_3$Sb$_5$ would naively have a higher $T_c$, consistent with reports from mesoscale devices that suppress CDW competition and report a higher $T_c$ \cite{PhysRevLett.127.237001, sun2025clean}. 

Doping to $5\%$ holes/f.u. suppresses the zero-resistance $T_c$ and broadens the transition again.  This correlates to the phase boundary where the CDW state switches to quasi-1D, incommensurate correlations.  These short-range correlations can be modeled as a single-\textbf{q} state in a three-domain model with an anisotropic correlation volume.  Interlayer correlations only extend over $\approx2$ unit cells with anti-phase correlations between neighboring layers; however charge correlations are substantially longer in-plane, yet still anisotropic. Correlations along \textbf{q} are short-range ($\xi_{\parallel}\approx$ 10 unit cells) whereas those orthogonal to \textbf{q} are longer ($\xi_{\perp}\approx$ 30 unit cells).  The resulting quasi-1D correlations map to an order whose correlations are strongest along the in-plane, nearest-neighbor V-V distance, suggesting a dimer instability.  

3D-$\Delta$PDF maps directly resolve V-V dimer formation and forward Monte Carlo modeling qualitatively capture the scattering data assuming short-range domains of one family of chains. While a 3\textbf{q} distortion is favored in models of the pristine compound, perturbations such as strain have shown that lengthened V-V dimer chains are a predicted, nearby instability \cite{zhang2024atomistic}.  A six-orbital Hubbard model has further resolved the presence of a nearby incommensurate single-\textbf{q} instability which can be stabilized via pressure (due to self-doping) \cite{PhysRevLett.127.177001, tazai2022mechanism}.  The chains of V-V in this model form under light electron-doping; however similar models exploring instabilities under hole-doping are not explored to the best of our knowledge.  


Generally, a single-\textbf{q}, L-point instability is allowed and predicted in large areas of the phenomenological phase diagram of AV$_3$Sb$_5$ compounds \cite{PhysRevB.104.214513, PhysRevB.107.205131}. However, the energy lowering due to displacements of an R=(3/8, 0, 1/2) point mode was not considered in prior \textit{ab initio} calculations. \textit{Ab initio} calculations shown in Fig. 4 (f-k) show that a single-\textbf{q} R-point displacement lowers the energy by an amount comparable to that of the L-mode, and, under small hole concentrations, the two phases become nearly degenerate \cite{Supplemental}. This is the likely origin of the shift in wave vector toward $\delta=3/8$, and it further suggests that the origin of the quasi-1D anisotropic correlation volume is due to a distribution of $\delta$ values arising from an interplay between this near-degeneracy and local disorder.  

Curiously, recent high-pressure studies show the emergence of a long-range ordered single-\textbf{q} instability as the parent 3\textbf{q} CDW state in CsV$_3$Sb$_5$ is suppressed \cite{PhysRevLett.133.236503}.  This long-range CDW state appears at a commensurate \textbf{q}=(3/8, 0, 1/2), and the new pattern of charge modulation corresponds to a minimum in the double-dome SC structure reported under pressure \cite{PhysRevLett.126.247001, yu2021unusual}.  

In both hole-doping and pressure-based studies, the boundary between the two SC domes corresponds to the onset of long-range single-\textbf{q} CDW order.  There are, however, differences between the two cases of hole-doping and pressure.  Pressure, naively induces electron self-doping, and the single-\textbf{q} order vanishes upon transitioning into the second dome. Regardless of the origin of these differences, short-range order of this new state and fluctuations between these two CDW states have a strong impact on the SC phase diagram of CsV$_3$Sb$_5$.



Future theoretical work resolving the microscopic origin of the quasi-1D V-V dimer correlations in hole-doped CsV$_3$Sb$_5$ is highly desired, in particular resolving the interplay between these dimer correlations and the superconducting phase.  The disappearance of these short-range ordered dimers coincides with the suppression of SC, suggesting the two instabilities are linked.  Our results establish a strong, experimentally demonstrated link between the properties of the high-temperature CDW correlations and the resulting SC properties in the $A$V$_3$Sb$_5$ class of kagome superconductors.   


This research was supported by the UC Santa Barbara NSF Quantum Foundry funded via the Q-AMASE-i program under award DMR-1906325. Work in the University of Minnesota was supported by the NSF CAREER grant DMR-2046020 and through the University of Minnesota MRSEC under Award Number DMR-2011401.  We acknowledge the use of shared facilities of the NSF MRSEC at UC Santa Barbara [DMR 1720256] and the Center for Scientific Computing, supported by the California Nano Systems Institute, the NSF MRSEC [DMR 1720256] and NSF CNS 1725797. The MRL Shared Experimental Facilities are supported by the MRSEC Program of the NSF under Award No. DMR 2308708; a member of the NSF-funded Materials Research Facilities Network (www.mrfn.org). This work is based on research conducted at the Center for High-Energy X-ray Sciences (CHEXS), which is supported by the National Science Foundation (BIO, ENG and MPS Directorates) under award DMR-2342336. Work by B.R.O. was supported by the U.S. Department of Energy (DOE), Office of Science, Basic Energy Sciences (BES), Materials Sciences and Engineering Division.

\bibliography{bibliography}

\end{document}